# Effects of "Stripes" on the Magnetic Excitation Spectra of $La_{1.48}Nd_{0.4}Sr_{0.12}CuO_4$

Masafumi Ito[1], Yukio Yasui[1], Satoshi Iikubo[1], Minoru Soda[1], Masatoshi Sato[1],
Akito Kobayashi[1] and Kazuhisa Kakurai[2]

[1]*Department of Physics, Division of Material Science, Nagoya University,
Furo-cho, Chikusa-ku, Nagoya 464-8602*
[2]*Advanced Science Research Center, JAERI, Tokai, Ibaraki 319-1195*

## Abstract

The wave vector($q$)- and energy($\omega$)-dependent magnetic excitation spectra $\chi''(q,\omega)$, have been studied in the temperature($T$) range of 10 K$\leq T <$150 K on single crystals of $La_{1.48}Nd_{0.4}Sr_{0.12}CuO_4$, which has the static "stripe" order in the low temperature phase. The incommensurability $\delta$ of the peak of $\chi''(q,\omega)$ found by the $q$-scans and the peak width $\Delta q$ exhibit characteristic $T$-dependences. Observed profiles show clear incommensurate structure even at rather high temperatures. The results indicate that effects of the slowly fluctuating "stripes" exist not only in the low temperature tetragonal (LTT) phase but also in the orthorhombic phase(LTO1) of the present system. The persistence of the local structural characteristics of the low-$T$ phase seems to be important.

corresponding author: M. Sato (e-mail: msato@b-lab.phys.nagoya-u.ac.jp)

KEYWORDS:    stripe, magnetic excitation spectra, neutron scattering

Measurements of the magnetic excitation spectra $\chi''(q, \omega)$ of high-$T_c$ superconductors are powerful methods to collect information on their electronic nature, where $q$ and $\omega$ are the wave vector and the energy of the excitation.[1-3]

The $T_c$-suppression found in the La214 systems such as $La_{2-x}Ba_xCuO_4$ and $La_{2-y-x}Nd_ySr_xCuO_4$ at around $x\sim1/8$(1/8 anomaly), is considered to be caused by the "stripe" ordering.[4,5] On the other hand, there exists a suggestion that dynamical "stripes" or the fluctuating "stripes" play an important role in the occurrence of high $T_c$ superconductivity.[6] To answer a question whether the "stripes" are really relevant to the mechanism of the high-$T_c$ superconductivity, a lot of studies have been carried out.

For a single crystal of $La_{2-y-x}Nd_ySr_xCuO_4$, superlattice peaks in $\chi''(q, \omega)$ were found at $q=(0.5\pm\delta,0.5)$ and $(0.5,0.5\pm\delta)$(in tetragonal notation).[7-9] If the slowly fluctuating "stripes" (SFS) or the dynamical "stripes" exist in other high-$T_c$ systems, similar incommensurate(IC) structure is also expected in $\chi''(q,\omega\neq 0)$. Then, one might consider that the IC peaks observed in superconducting YBCO crystals at the $q$ points similar to those found in 1/8-doped $La_{2-y-x}Nd_ySr_xCuO_4$[10,11] can be interpreted as the experimental evidence for an effect of dynamical "stripes".

To clarify if the IC peaks observed in $\chi''(q, \omega)$ of YBCO are really relevant to the SFS, we have tried to fit of the calculated $\chi''(q, \omega)$ of $YBCO_y$ ($y\geq 6.5$) to the observed data.[12,13] In the calculation, an expression of the dynamical spin susceptibility $\chi(q,\omega) =\chi^0(q, \omega)/\{1+J(q)\chi^0(q, \omega)\}$,[14,15] where $\chi^0(q, \omega)$ is the Lindhard function of the electron system without the exchange coupling $J(q)$ between the Cu-Cu electrons was used. We chose the effective band parameters $(t_0,t_1,t_2)$ to reproduce the Fermi surface. Consideration of the energy broadening($\Gamma$) of the quasi particles was found to suppress the antiferromagnetic ordering. For the $d$-wave gap parameter, qualitative characteristics of all the $q$-, $\omega$- and $T$-dependences of $\chi''(q, \omega)$ were well reproduced. Then, the existence of "stripes" is not necessarily required to explain the observed $\chi''(q,\omega)$.

In the present work, we have deliberately chosen $La_{1.48}Nd_{0.4}Sr_{0.12}CuO_4$ which has the static "stripe" order below $T_e$ and measured the magnetic excitation spectra both in the $T$-regions above and below $T_e$ to clarify how effects of static and dynamical "stripes" appear on $\chi''(q,\omega)$.

Single crystals grown by a traveling solvent floating zone method, were annealed under oxygen atmosphere for 50 hours. In the experiments, six aligned crystals with the typical size of about 7 mm$\phi$ ×30 mm were used. In the neutron measurement, the triple axis spectrometer ISSP-PONTA installed at JRR-3M of JAERI in Tokai was used. Scattered neutron energy ($E_f$) was selected at 14.7 meV using the 002 reflection of Pyrolytic graphite (PG) both for the monochromator and the analyzer. A PG filter was placed after the sample to eliminate the higher order contamination from the scattered beams. The horizontal collimations were 40'-40'-80'-80'. The sample crystals were oriented with the [001] axis vertical. They were mounted in an Al can with He exchange gas which was attached to the cold finger of a Displex type closed cycle refrigerator.

The reflections corresponding to the spin ordering, the lattice distortion induced by the charge ("stripe") order and the 100/010 nuclear reflections which indicate the occurrence of the transition from the low temperature orthorhombic(LTO1) phase with the space group B*mab* to the low temperature tetragonal(LTT) phase via an intervening orthorhombic(LTO2) phase with the space group P*ccn*, are measured along the lines shown by the arrows (a)-(c) in the inset of Fig. 1(a), respectively, and the peak intensities of the reflections observed as functions of $T$ at the scattering vector $\boldsymbol{Q}=(2+2\delta,0)$, $(0.5,0.5-\delta)$ and $(1,0)$ are shown in Figs. 1(a)-1(c), respectively. In these figures the background counts are already subtracted. A typical profile of the reflection corresponding to the spin ordering is shown in the inset of Fig. 1(b), from which $\delta(\omega=0)=0.119\pm0.0005$ is obtained. We can see in Fig.1(c), the structural transitions from the LTO1→LTO2 → LTT start at temperature $T_s$ (~69K) with decreasing $T$. Due to the existence of the intervening LTO2 phase, the intensity does not change discontinuously but continuously in the narrow region below $T_s$. The superlattice peaks at $\boldsymbol{Q}=(2+2\delta,0)$ and $(0.5,0.5-\delta)$ which are considered to be induced by the charge and magnetic orderings, respectively, appear at almost equal temperatures, which we call $T_e$. Although the relation $T_e \cong T_s$ holds, $T_e$ is considered as the temperature of the second order transition in the LTT (or LTO2) phase, because the order parameter of the "stripes" grows with decreasing $T$ from zero continuously. In other words, even if the LTT (or LTO2) phase persisted up to higher temperature, the "stripe" order disappeared at $T_e$ with increasing $T$. Fujita et al.[16] have suggested that the static "stripes" are stabilized in the LTT/LTO2 phase, and that "stripe" order is suppressed just above $T_s$ in $La_{2-y-x}Nd_ySr_xCuO_4$. However, the gradual growth of the superlattice peak in Fig. 1(a) indicates that such the suppression may not be appropriate for this system.

Data of $\chi"(\boldsymbol{q},\omega)$ have been collected by scanning the scattering vector $\boldsymbol{q}$ in $k$ direction (parallel to the $\boldsymbol{b^*}$ direction) around $(0.5,0.5)$ at several temperatures between 7 K and 150 K with the transfer energy being fixed at 2.5, 5.0, 8.0 and 12.0 meV. Detailed $T$-dependences of $\chi"(\boldsymbol{q},\omega)$ have been taken at 2.5 and 5.0 meV. Typical profiles at $\omega=2.5$ meV are shown in Fig. 2. Below $T_e$, the sharp excitations have been observed at $(0.5,0.5\pm\delta)$. The solid lines show the results of the fittings with the double Gaussian line, where used parameters are $\delta$, the peak width $\Delta\boldsymbol{q}$, the scale factor and the background counts which are assumed to be linear in $\boldsymbol{q}$. By the fittings, the values of $\delta$ in low temperature region are found to be about 0.118, which is consistent with that reported in ref. 8. The IC structure of $\chi"(\boldsymbol{q},\omega)$ persists up to high temperatures and the $\delta$ values are determined to be ~0.105 and ~0.102 at 75 K and 92 K, respectively, by the fittings(solid lines). The broken lines in Fig. 2 show, for comparison, results of the fittings with $\delta$ being fixed at the low temperature value of 0.118. The much worse fittings than the solid lines indicate that the $\delta$ values above $T_s$ are certainly smaller than those below $T_s$.

The profiles at $\omega$=5.0 meV are shown in Fig. 3. For $T<T_s$, sharp excitation peaks are observed at (0.5,0.5±$\delta$), where $\delta\cong$0.116 is obtained by the fittings (solid lines). The IC structure persists up to the highest temperature(143 K) studied here. The $\delta$ values at 92 K and 122 K are determined to be 0.091 and 0.090, respectively. The broken lines show results of fittings with fixed $\delta$ of =0.116. As in the case of $\omega$=2.5 meV, the values of $\delta$ above $T_s$ are clearly smaller than those observed below $T_s$.

Figure 4 shows the $T$-dependence of $\delta$. Filled and open circles correspond to $\omega$=2.5 and 5.0 meV, respectively. Similar characteristics have been observed for both energies: At low temperatures, $\delta$ is almost $T$-independent and it begins to decrease with increasing $T$ in the $T$-region around $T_s \cong T_e$. The $\delta$ values at low temperatures for $\omega$=2.5 and 5.0 meV are close to $\delta(\omega=0)$ but a slight decrease of $\delta$ can be seen with $\omega$.

In Fig. 5, the temperature dependence of the peak width $\Delta\boldsymbol{q}$ of $\chi"(\boldsymbol{q},\omega)$ is shown, where the resolution correction is already made. Filled and open circles correspond to $\omega$=2.5 and 5.0 meV, respectively. The $\Delta\boldsymbol{q}$ values are very small at low temperatures and exhibit significant increase with increasing $T$ through $T_s$.

Figure 6 shows the temperature dependence of the $\boldsymbol{q}$-integrated intensities ($S(\omega)$) of the magnetic excitations. The rate of the intensity decrease with decreasing $T$ seems to change for both energies of $\omega$=2.5 and 5.0 meV at around the temperature where the magnetic reflection peak shown in Fig. 1(b) appears. We have not observed significant changes of the integrated intensities of $S(\omega)$ above this temperature.

Now, we argue effects of "stripes" on $\chi"(\boldsymbol{q},\omega)$. We have applied a model similar to that used for YBCO$_y$,[12,13)] which does not consider effects of "stripes". For the present system, the effective band parameter $t_0$ is chosen to be -40 meV and $t_1$ and $t_2$ are chosen to be -$t_0$/6 and 0, respectively to reproduce the 2-dimensional Fermi surface shape of La$_{2-x}$Sr$_x$CuO$_4$.[17)] The parameter $J_0$ used in the expression $J(\boldsymbol{q})=J_0(\cos q_x a + \cos q_y a)$ with $a$ being the Cu-Cu lattice spacing is chosen to be 50 meV and the chemical potential $\mu$ is so chosen to reproduce the $\delta$-value at $T_s$. We just set the superconducting gap to be zero(The $\delta$ value is slightly dependent on the gap magnitude.). A small value of $\Gamma$ of 2 meV (full width at half maximum) is also used. Other details can be found in refs. 12 and 13.

For these parameters, the $T$-dependence of $\delta$ above $T_s$ can essentially be reproduced, as shown in Fig. 4. (Antiferromagnetic ordering does not take place for the above set of parameters.) However, it has been found not to be able to obtain the profile widths as sharp as the observed ones even for the present small value of $\Gamma$. The $T$-insensitive behavior of $\delta$ below 50 K has not been reproduced in the present calculation which does not consider the "stripe" ordering. The low temperature values of $\delta(\omega\neq0)$ are close to those of $\delta(\omega=0)$=0.119, suggesting that the $\delta$ value below $T_s$ is determined by the periodicity of the static "stripes".

We discuss if the existence or non-existence of the SFS is related with the structure of the system. In the analyses described above, the rather sharp widths of the IC peaks in the high temperature region can hardly be reproduced even by using the very small $\Gamma$. (Judging from the transport and optical studies,[18,19] $\Gamma$ is expected to be as large as that of YBCO$_y$.) This result has a clear contrast to the case of YBCO$_y$, where detailed characteristics of $\chi"(q,\omega)$ can be reproduced up to $\omega\sim 25$ meV by using the $\Gamma$ value as large as the experimentally observed ones.[20,21] It may be consistent with an idea that there exists a kind of spatial separation of the electron system into hole-rich and hole-poor parts, or the existence of the "stripes", because the hole-poor parts mainly determine the width of the peaks. The observed decrease of $\Delta q$ in rather wide $T$-region as $T$ approaches $T_s$ from above, can be considered to be due to the growth of the "stripe" fluctuations, which directly indicates the existence of the SFS even above $T_e$ (or $T_s$).

Based on the above arguments, we think that in the LTO1 phases of La$_{2-y-x}$Nd$_y$Sr$_x$CuO$_4$, the SFS exist. The characteristic frequency $\omega$ of the fluctuations should be smaller than 5 meV in the present temperature region. (Note that in YBCO$_y$, even for $\omega$ as large as ~25 meV, effects of the "stripes" have not been observed.) Here, it should be noted, however, that recent work by Han *et al.* showed that the local structure of the present system is not the ideal LTO1 structure but like the LTO2 one even at room temperature.[22] This structural characteristic may induce the short range "stripe" correlation even above $T_s$. The persistence of the local structural characteristics of the low temperature phase[22] may play an essential role in realizing the SFS in the LTO1 phase, though it is not easy to distinguish if the SFS can exist even in the CuO$_2$ planes with the ideal LTO1 structure. We note here that in other La214 systems, the local structure has similar characteristics to those of the present system, too.[23,24]

Finally, we briefly comment on the reported data of $\chi"(q,\omega)$ of La$_{2-x}$Sr$_x$NiO$_4$ with $x$=0.275.[17] Although $\delta$ of the system seems to be $T$-dependent, only the very small number of the temperature points at which $\chi"(q,\omega)$ was measured makes it difficult to argue the $T$-dependence of $\delta$ in detail. It should be also noted that $\chi"(q,\omega)$ of the system may not properly be treated by the model used in the above calculation, and therefore it is not appropriate to compare the experimental results with those of the present system which has itinerant electrons at least above $T_e$.

In summary, the magnetic excitation spectra $\chi"(q,\omega)$ of La$_{1.48}$Nd$_{0.4}$Sr$_{0.12}$CuO$_4$ up to rather high temperature have been reported. The $T$-insensitive behavior of $\delta$ below 50 K and the persistence of the rather clear IC structure is one of the characteristics of the observed results. It has been found that the $T$-dependence of $\delta$ can be reproduced by the simple model calculation based on the expression of the generalized susceptibility $\chi(q,\omega) = \chi^0(q,\omega)/\{1+J(q)\chi^0(q,\omega)\}$. The arguments on the relationship between the observed results and the "stripe" fluctuations have been presented.

Acknowledgment-Two of the authors(M. I and S. I) are supported by a Research Fellowship of the Japan Society for the Promotion of Science for Young Scientists.

Figure captions

Fig. 1  Temperature dependence of the superlattice peaks. The top, middle and bottom figures correspond to the reflections originating from the charge, magnetic orderings and LTO1→LTO2 transition, respectively. The intensities for all the reflections are scaled by the values at the lowest temperatures. In the inset of (a), the 2D reciprocal space is shown. In the inset of (b), a profile of the superlattice peak originating from the spin ordering is shown. See text for details.

Fig. 2  Neutron scattering profiles of the magnetic excitations taken at the energy of 2.5 meV at several temperatures. Solid lines represent results of the profile fittings with double Gaussian line. Broken lines represent results of profile fittings with double Gaussian line with fixed δ of 0.118 obtained at low temperature (~10 K). The zeros of the vertical axis are shifted upwards by 150, 330, 530 counts/19500 kmon. for $T$=75, 40 and 10 K, respectively.

Fig. 3  Neutron scattering intensities of the magnetic excitations taken at the energy of 5.0 meV at several temperatures. Solid lines represent results of the profile fittings with double Gaussian line. Broken lines represent the results of profile fittings with fixed δ of 0.116 obtained at low temperature (~10 K). The zeros of the vertical axis are shifted upwards by 220, 350, 500 counts/19500 kmon. for $T$=92, 33 and 10 K, respectively.

Fig. 4  Temperature dependences of δ of the magnetic excitations. Filled and open circles represent the data points for ω=2.5 meV and 5.0 meV, respectively. Solid line shows the result of the calculation described in the text(ω=2.5 meV). It deviates in the $T$-region below $T_e$.

Fig. 5  Temperature dependences of the peak width Δ$q$ (FWHM) of the magnetic excitation spectra taken along the $k$ direction (see Fig. 1(a)) at ω=2.5 and 5.0 meV. Filled and open circles represent the data points for ω=2.5 meV and 5.0 meV, respectively. Solid line is a guide to the eye.

Fig. 6  Temperature dependence of the integrated intensities ($S(\omega)$) of the magnetic excitation peaks at ω=2.5 meV (top) and 5.0 meV (bottom).

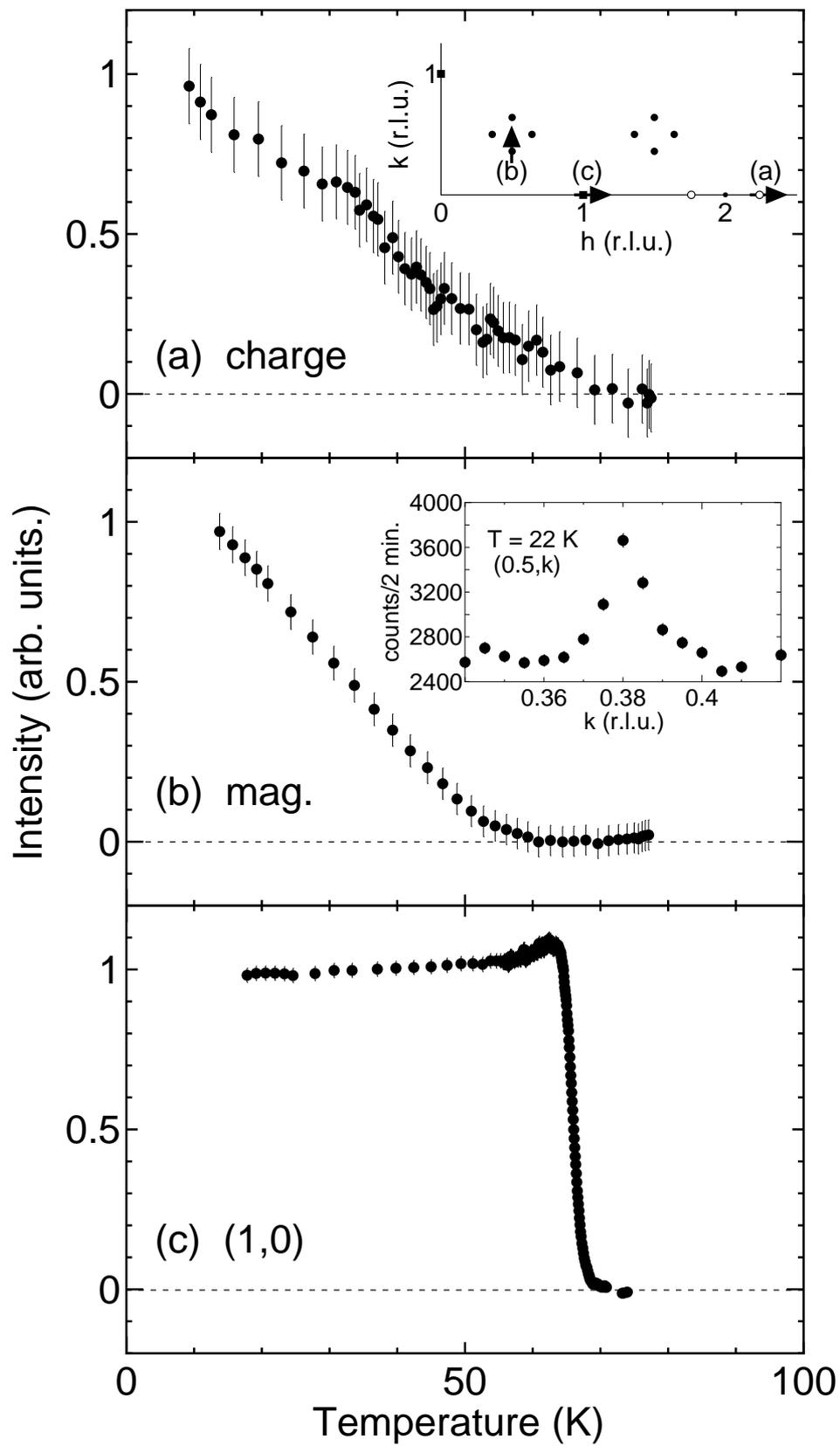

Fig. 1
M. Ito et al.

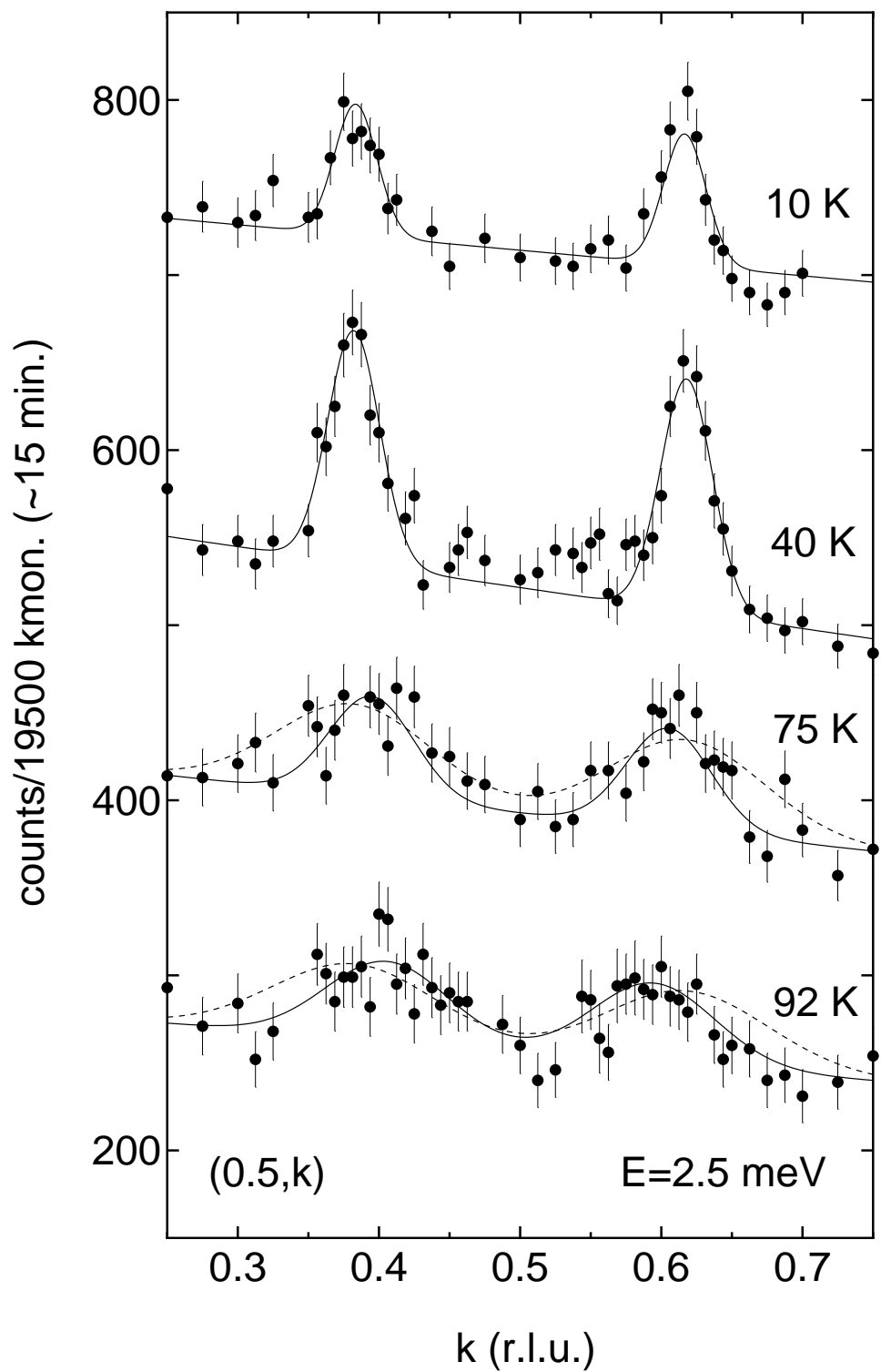

Fig. 2
M. Ito et al.

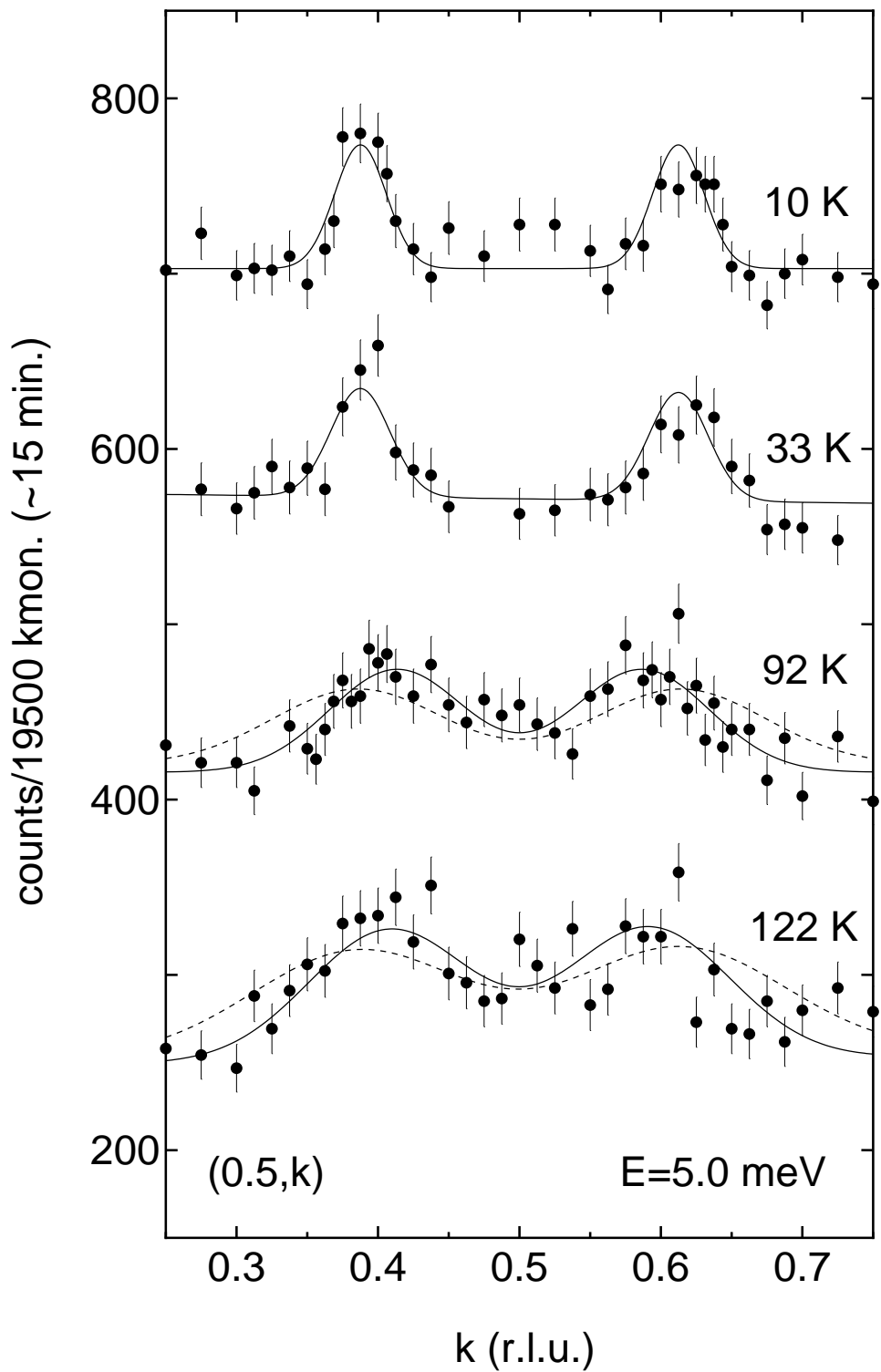

Fig. 3
M. Ito et al.

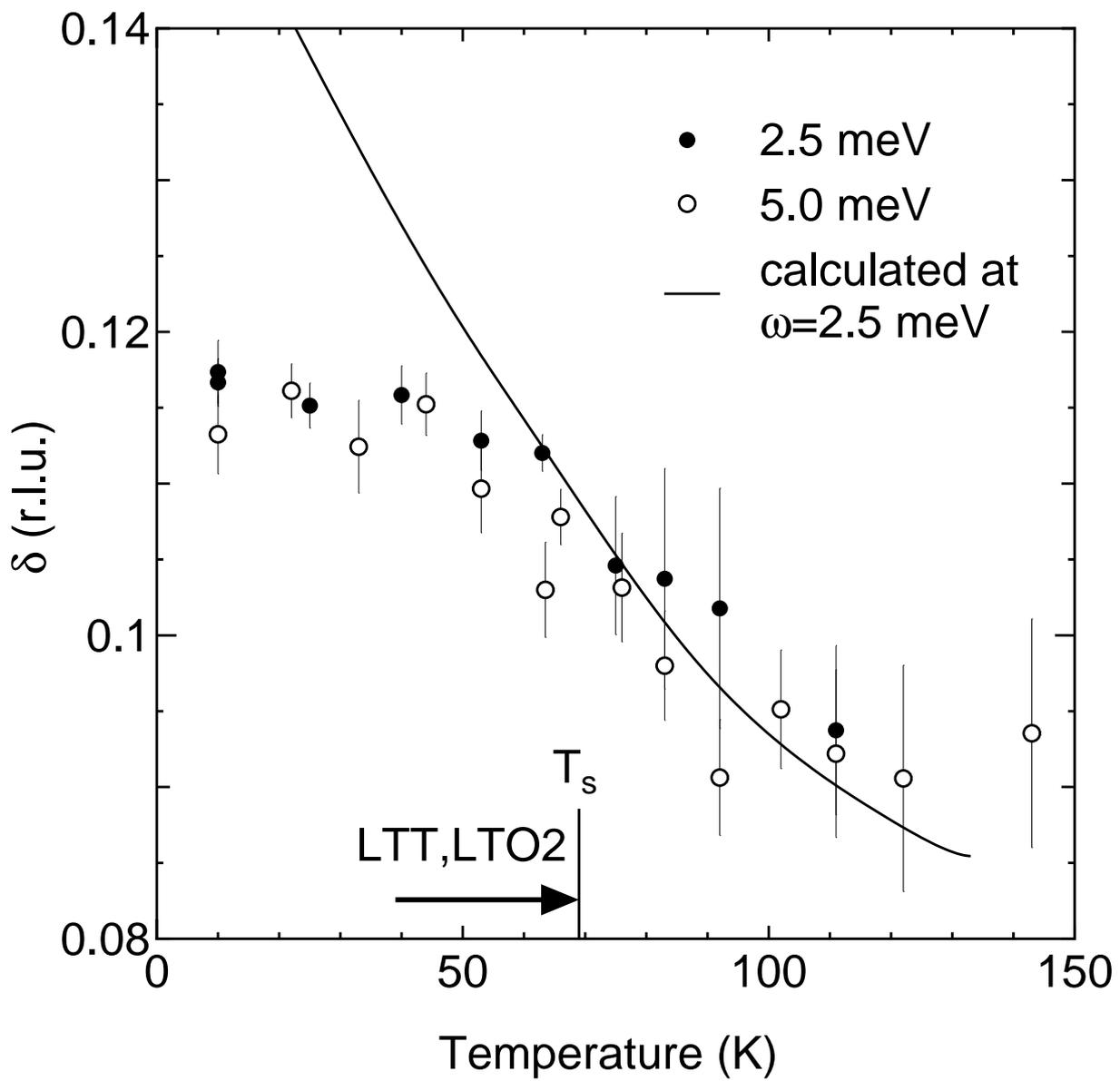

Fig. 4
M. Ito et al.

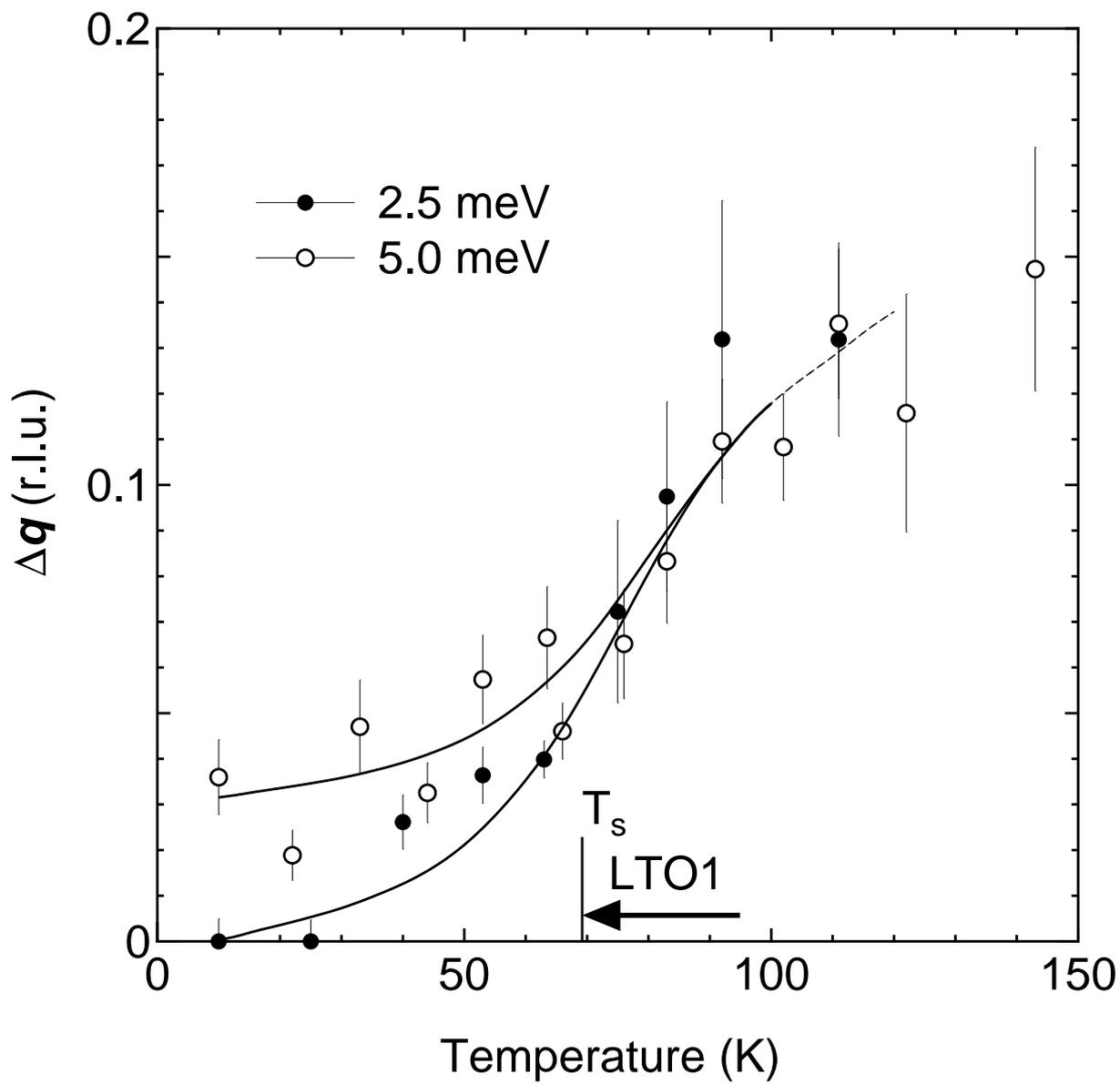

Fig. 5
M. Ito et al.

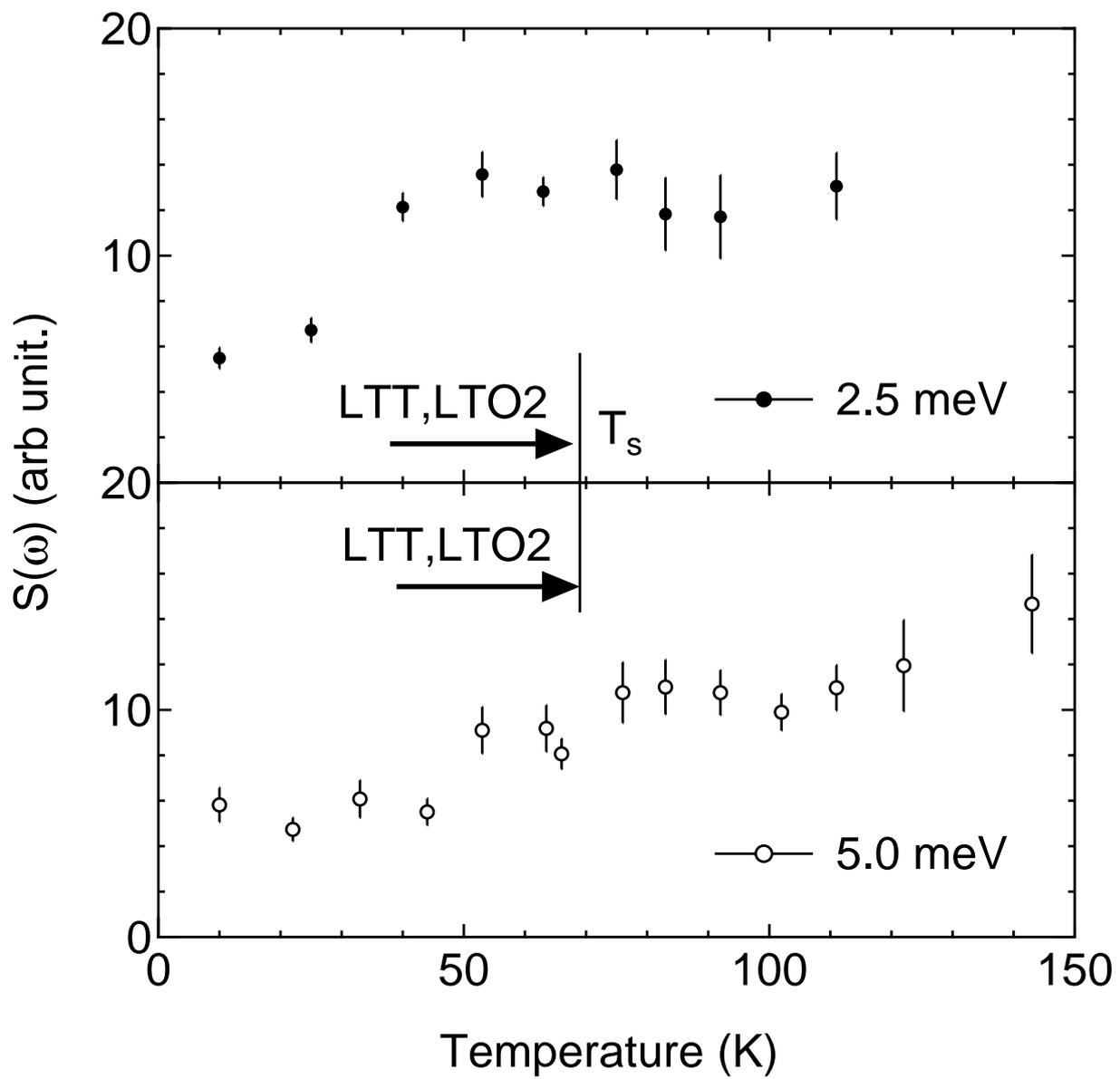

Fig. 6
M. Ito et al.